\documentstyle[aps,psfig,multicol]{revtex}
\begin{document} 
\draft
\title{Breakdown of the Onsager reaction field theory in two dimensions}
\author{Avinash Singh}
\address{Department of Physics, Indian Institute of Technology, Kanpur-208016, India}
\maketitle
\begin{abstract} 
It is shown that for the spin 1/2 anisotropic Heisenberg 
model the result for the transition temperature
is completely insensitive to the anisotropy
within the Onsager reaction field theory,
which yields a vanishing $T_c$ in
two dimensions, in total conflict with the expected 
finite $T_c$, increasing with anisotropy.
This establishes that the breakdown of the Onsager reaction field
theory in two dimensions is not limited to the Ising
model, but actually extends over the whole range of anisotropy 
$0 \le |J_{ij}^x /J_{ij}^z|\; ,  
|J_{ij}^y /J_{ij}^z| < 1$. 
Therefore, for the isotropic case, 
the related results of vanishing $T_c$ and exponential
dependence of spin correlation length
should be seen as mere coincidence.

\end{abstract} 
\begin{multicols}{2}\narrowtext

The reaction field theory first applied by Onsager
to the many dipole problem,\cite{onsager} and subsequently 
extended to interacting spin systems,\cite{brout.thomas} 
involves an improvement over the local mean field theory 
by subtracting from the orienting field a reaction field,
involving that part of the polarization which is due to 
correlations. 
Thus, in the context of spin systems, 
the effective orienting field on a given spin 
excludes that part of the spin polarization of surrounding spins
which is due to their correlation with that spin.
It was also found\cite{brout.thomas} that ``the Onsager 
prescription of subtracting out the reaction field is
the necessary modification of molecular mean field
theory in order to guarantee the fluctuation-response function 
relation of statistical mechanics, i.e., susceptibility = fluctuation.''
The inclusion of the reaction field introduces a degree of freedom 
which allows for self consistency between susceptibility and correlation. 

Viewed in terms of an improvement over mean field theory, 
the reduction in the magnetic transition temperature $T_{\rm c}$,
relative to the mean field transition temperature $T_{\rm c}^{\rm MF}$, 
has been discussed for the isotropic Heisenberg model
in three dimensions.\cite{white} 
While the original work on the Ising model\cite{brout.thomas}
does not discuss the transition temperature, 
formally the treatment is identical
to that for the Heisenberg model, resulting in an {\em identical}
final expression for the correction in $T_{\rm c}$:
\begin{equation}
\frac{T_{\rm c}}{T_{\rm c}^{\rm MF}} =
\left [\sum_{\bf q} \frac{1}{1-J({\bf q})/J({\bf Q})} \right ]^{-1}.
\end{equation}
Here $ J({\bf q})=\sum_j J_{ij} e^{i{\bf q}.{\bf r}_{ij}}$ is the Fourier transform of the
spin coupling $J_{ij}$ between spins at sites $i$ and $j$,
and ${\bf Q}$ is the ordering wavevector.

The implications of this 
identical result for the Heisenberg and the Ising model,
with regard to the continuous spin-rotation  symmetry, 
have not been discussed in the literature so far.
This is especially pertinent in two dimensions,
where the ${\bf q}$ sum in Eq. (1) diverges logarithmically
for short-range spin interactions,
yielding a vanishing transition temperature. 
Hence, for the Ising model there is obviously 
a breakdown of the Onsager method in two dimensions,
as even in the absence of a continuous spin-rotation symmetry,
a vanishing $T_{\rm c}$ is obtained,
indicating no connection between the Onsager theory and the
Mermin-Wagner theorem. 

In view of this breakdown for the Ising model,
is the Onsager-theory result of vanishing $T_{\rm c}$ 
for the two-dimensional Heisenberg model, then, 
in accordance with the Mermin-Wagner theorem ?
In other words, is the vanishing $T_{\rm c}$ result
crucially dependent on the
continuous spin-rotation symmetry, or is it a mere coincidence ? 
This is the question we will explore
in this paper. For this purpose, we consider an
anisotropic spin 1/2
Heisenberg model with anisotropy in spin space. 
With the removal of the continuous spin-rotation symmetry, 
if a non-vanishing $T_{\rm c}$ is obtained, then indeed the
Onsager method is sensitive to the continuous symmetry.
However, if the vanishing $T_{\rm c}$ result persists,
then this will confirm that the continuous symmetry has no role
to play, and the vanishing $T_{\rm c}$ is a coincidence,
indicating a breakdown of the
Onsager reaction field method in two dimensions. 

These questions are relevant in view of the recent 
use of the Onsager method in quantitative studies of
the N\'{e}el temperature $T_{\rm N}$, and the spin correlation length $\xi$,
for the spin 1/2 Heisenberg antiferromagent in two and 
three dimensions.\cite{logan}
In addition to the vanishing $T_{\rm N}$, an exponentially large
spin correlation length at low temperatures was also obtained
in two dimensions. 
This exponential dependence arises from the same logarithmic
divergence in Eq. (1) which leads to the vanishing $T_{\rm N}$,
hence the two results are not independent, but follow
from the same source.

We find, not surprisingly,  
that the introduction of the anisotropy in spin space 
does not affect the vanishing $T_{\rm c}$ result within Onsager theory,
indicating a complete {\em insensitivity} to the continuous symmetry.
Whereas, an anisotropy driven gap in the spectrum of 
transverse spin fluctuation modes (magnons), 
and a non-vanishing $T_c$ of the order of the
gap are characteristic features of the anisotropic Heisenberg model.
The inescapable conclusion then is that the Onsager method breaks down
in two dimensions, and the results of vanishing $T_{\rm N}$ and exponential
spin correlation length for the Heisenberg model, following from Eq. (1), are coincidental. 

\section{Anisotropic Heisenberg model --- Onsager Correction}
We consider the spin 1/2 anisotropic Heisenberg model with anisotropy in spin space,
\begin{equation}
H=-\sum_{\langle ij\rangle}
[J_{ij} ^x S_{ix} S_{jx} +
J_{ij} ^y S_{iy} S_{jy} +
J_{ij} ^z S_{iz} S_{jz}] \; ,
\end{equation}
where $J_{ij} ^x = J_{ij} ^y$ and $|J_{ij} ^z|> |J_{ij} ^x|,|J_{ij} ^y|$, 
so that ordering first occurs, with decreasing temperature, 
in the z-direction. 
For an antiferromagnet the spin couplings are all negative. 
Considering the spin-symmetric, paramagnetic state at
$T>T_{\rm c}$, 
in the presence of an infinitesimal external magnetic field 
in the z-direction, the Weiss mean field,
\begin{equation}
H_{iz}^{\rm Weiss}=
H_{iz}^{\rm ext} + \sum_j J_{ij} ^z \langle S_{jz} \rangle \; ,
\end{equation}
acts as an effective orienting field on the spin
at site $i$ within the Weiss mean-field theory,
and the resulting polarization $\langle S_{iz}\rangle$
is self-consistently determined from 
\begin{equation}
\langle S_{iz} \rangle=
\sum_{\{S_{iz}\}} S_{iz} e^{\beta S_{iz}
H_{iz}^{\rm Weiss} } /
\sum_{\{S_{iz}\}} e^{\beta S_{iz} H_{iz}^{\rm Weiss} } \; .
\end{equation}

In the limit of infinitesimal field, upto linear terms
\begin{eqnarray}
\langle S_{iz} \rangle &=&
[\sum_{\{S_{iz}\}} S_{iz}^2 /(2S+1) ] . 
\beta H_{iz}^{\rm Weiss} \nonumber \\
&=&
\langle S_{iz}^2 \rangle _{\rm free\; spin} . 
\beta H_{iz}^{\rm Weiss} \nonumber \\
&=& 
\chi^0 . H_{iz}^{\rm Weiss}  \; ,
\end{eqnarray}
where 
\begin{equation}
\chi^0 \equiv \beta 
\langle S_{iz}^2 \rangle _{\rm free\; spin} = \beta S(S+1)/3 \; .
\end{equation}
This result is actually valid for any spin quantum
number $S$, for which $S_{iz}$ takes all integer-step 
values between $-S$ and $S$.

From Eqs. (5) and (3), we obtain after Fourier transformation, 
\begin{eqnarray}
\langle S_z({\bf q}) \rangle 
&=& 
\sum_i \langle S_{iz} \rangle . e^{i{\bf q}.{\bf r}_i}  \nonumber \\
&=& 
\chi^0  
\sum_i [H_{iz}^{\rm ext} + \sum_j J_{ij} ^z \langle S_{jz} \rangle ]
e^{i{\bf q}.{\bf r}_i} \nonumber \\
&=& \chi^0 [H_z^{\rm ext}({\bf q}) + J_z ({\bf q})\langle S_z({\bf q})
 \rangle ] \; ,
\end{eqnarray}
where $J_{ij}^z =\sum_{\bf q'} J_z ({\bf q'}) \exp{-i{\bf q'}.
({\bf r}_i-{\bf r}_j)}$.
This yields the RPA susceptibility as the magnetization response
to an infinitesimal external field,
\begin{eqnarray}
\chi_z ({\bf q}) 
&=& 
\langle S_z({\bf q}) \rangle /H_z^{\rm ext}({\bf q}) \nonumber \\
&=& 
\frac{\chi^0}{1-\chi^0 J_z ({\bf q})} \; .
\end{eqnarray}

The mean field transition temperature $T_{\rm c}^{\rm MF}$ 
is obtained from the divergence of the spin susceptibility
at the ordering wavevector ${\bf Q}$, and from Eqs. (8) and (6),
\begin{equation}
T_{\rm c}^{\rm MF}=
J_z ({\bf Q}). \langle S_{iz}^2 \rangle _{\rm free\; spin} 
=J_z ({\bf Q}) S(S+1)/3 \; ,
\end{equation}
which is valid for general spin $S$, and exhibits
no effect of the anisotropy because the analysis involves
an isolated  single spin in the frozen mean field.

Going over now to the Onsager reaction field correction, 
an improvement over the mean-field approximation is obtained
by subtracting a term from the mean field which
is proportional to the mean field itself:
\begin{equation}
H_{iz}^{\rm eff}= \sum_j J_{ij} ^z \langle S_{jz} \rangle 
- \sum_j \lambda_{ij} J_{ij} ^z \langle S_{iz} \rangle \; ,
\end{equation}
where the parameters $\lambda_{ij}$ are temperature dependent,
and characterize the Onsager reaction field. 
$ \lambda_{ij} $ is related to the correlation 
$\langle S_{iz} S_{jz} \rangle$, 
and the subtracted polarization 
$ \lambda_{ij} \langle S_{iz} \rangle$ is that part 
of the polarization $\langle S_{jz} \rangle$ 
which is due to correlation with the spin
at site $i$. 

Following the earlier analysis for the spin susceptibility
leading to Eq. (8), one obtains
\begin{equation}
\chi_z ({\bf q}) 
= \frac{\chi^0}{1-\chi^0 (J_z ({\bf q})-\lambda)}  \; ,
\end{equation}
where $\lambda =\sum_j \lambda_{ij} J_{ij}^z$.
As before, $T_c$ is obtained
from the divergence in the spin susceptibility at the
ordering wavevector ${\bf Q}$, and one obtains
\begin{equation}
\frac{T_{\rm c}}{T_{\rm c}^{\rm MF}}=
[1-\lambda(T_{\rm c})/J_z(Q)] \; .
\end{equation}
The reaction field reduces the net mean field, and hence
the transition temperature. 
This shows that an important characteristic of the 
Onsager correction is that it acts {\em within} the 
mean field theory. This is in contrast to
the spin-wave theory which involves transverse 
spin-fluctuation corrections  
{\em about} the mean-field state.\cite{neel}

The presence of the $\lambda$ term in the susceptibility
expression in Eq. (11) 
introduces a degree of freedom which allows for 
self-consistency between the susceptibility and correlation.
The fluctuation-dissipation theorem provides the following connection
between the zero-field spin correlations and the spin 
susceptibility:\cite{white}
\begin{equation}
\beta \langle S_{iz} S_{jz} \rangle^0=
\sum_{\bf q} \chi_z ({\bf q}) e^{i{\bf q}.{\bf r}_{ij}} \; .
\end{equation}

Now, for the isotropic model in the paramagnetic state at $T>T_c$,
all integer-step spin states 
for $S_{iz}$ between $-S$ and $S$ are equally probable, 
and therefore for zero external field
$\langle S_{iz}^2 \rangle ^0 = S(S+1)/3$
for general spin $S$.
For the anisotropic model higher $|S_{iz}|$ values are
more likely due to the anisotropy, and therefore 
$\langle S_{iz}^2 \rangle ^0 = \alpha. S(S+1)/3$,
where $\alpha \ge 1 $ is a temperature-dependent factor
which accounts for the anisotropy. 

However, for the spin 1/2 case $\alpha=1$,
as the relation 
$\langle S_{iz}^2 \rangle ^0 = S(S+1)/3$
(with $S=1/2$)  continues to hold even in the presence of
the uniaxial anisotropy, 
because both spin states $S_{iz}= \pm 1/2$
are symmetrical and therefore 
equally likely in the paramagnetic state. 
This also implies that 
$\langle S_{iz}^2 \rangle ^0
=\langle S_{iz}^2 \rangle _{\rm free\; spin}$.
It is this equality which is responsible for identical
final expressions for the Onsager-corrected $T_c$ for
the Ising, Heisenberg and anisotropic Heisenberg models.
Therefore,  from Eq. (6),
\begin{equation}
\beta  \langle S_{iz}^2 \rangle ^0 
= \beta \alpha S(S+1)/3=\alpha \chi^0 \; ,
\end{equation}
which leads to the sum rule:
\begin{equation}
\beta \langle S_{iz}^2 \rangle^0=
\alpha \chi^0 = \sum_{\bf q} \chi_z ({\bf q}) 
= \sum_{\bf q}  \frac{\chi^0}{1-\chi^0 (J_z ({\bf q})-\lambda)} \; .
\end{equation}
Choosing $\lambda$ such that the above sum rule is satisfied 
ensures the self-consistency between the susceptibility
and the correlation determined from it. 
Taking the limit $T\rightarrow T_{\rm c}^{+}$
in Eq. (15),
using $\chi^0(T_{\rm c})J_z ({\bf Q})=
T_{\rm c}^{\rm MF}/T_{\rm c}$
from Eqs. (6) and (9), 
and eliminating $\lambda$ from the above equation and from
Eq. (12) determining $T_{\rm c}$, we obtain after 
straightforward algebra the following expression for
the corrected transition temperature,
\begin{equation}
\frac{T_{\rm c}}{T_{\rm c}^{\rm MF}} = \alpha . 
\left [\sum_{\bf q} \frac{1}{1-J_z({\bf q})/J_z({\bf Q})} \right ]^{-1}.
\end{equation}

For the $S=1/2$ anisotropic model $\alpha=1$, as discussed 
earlier, so that the result for the transition temperature
is totally insensitive to the anisotropy. 
The (logarithmic) divergence in two dimensions,
and therefore the vanishing $T_c$ result, 
remain unchanged even with the anisotropy.
This is in total conflict with the expected result
for the anisotropic model of a finite $T_c$,
increasing with anisotropy.
This establishes that the breakdown of the Onsager reaction field
theory in two dimensions is not limited to the Ising
model, but also holds for the anisotropic Heisenberg  model
in the whole range of anisotropy 
$0 \le |J_{ij}^x /J_{ij}^z| < 1$. 
Therefore, the vanishing $T_c$ result for the isotropic case
should be seen as a mere coincidence.

It is instructive to examine the origin of the divergence. 
The divergence is at the wavevector ${\bf q}={\bf Q}$, 
and follows from the definition of
the transition temperature where the 
susceptibility must diverge. 
Thus (not surprisingly) it is present for the Ising model,
as well as for the anisotropic Heisenberg model,
and has nothing to do with the presence of any 
soft Goldstone mode connected with the 
continuous spin-rotation symmetry.

Continuing with the spin 1/2 case, for which the sum-rule simplifies
to $\sum_q \chi_z({\bf q}) =\chi^0$, from Eqs. (13) and (15) one obtains:
\begin{equation}
\lambda_{ij}=\frac{\langle S_{iz} S_{jz} \rangle^0}
{S(S+1)/3} \; ,
\end{equation}
confirming that the Onsager parameters are related to the
spin correlations, and providing for a self-consistent determination 
of the correlation function 
$\langle S_{iz} S_{jz} \rangle^0$ in Eq. (13). 

For vanishing $T_c$, it follows from Eq. (12) that
$\lambda= J_z({\bf Q})$, and hence $|\lambda_{ij}|=1$,
indicating perfect NN spin correlations at $T=0$.
Thus $T_c$ vanishes simply because the
reaction field exactly cancels the mean field term in  
Eq. (10), reducing the effective orienting field to zero. 
This is a very different mechanism from that 
involved in a two-dimensional 
system possessing continuous spin-rotational symmetry,
where the vanishing of $T_c$
is associated with the long-wavelength
Goldstone modes.

In conclusion, it has been shown that for
the anisotropic $S=1/2$ Heisenberg  model
the Onsager reaction field theory yields a vanishing 
$T_c$ in two dimensions in the whole range of anisotropy 
$0 \le |J_{ij}^x /J_{ij}^z|\; ,  
|J_{ij}^y /J_{ij}^z| < 1$,
and generally shows a total insensitivity to the anisotropy factor.
This result is in total conflict with the expected 
finite $T_c$, increasing with anisotropy,
and confirms that the breakdown of the Onsager reaction field
theory in two dimensions is not limited to the Ising
model, but actually extends over the whole range of anisotropy.
Therefore, for the isotropic case, 
the related results of vanishing $T_c$ and exponentially large 
spin correlation length should be seen as a mere coincidence, 
unrelated to the Goldstone mode associated with the continuous 
spin-rotational symmetry. 

The Onsager reaction field theory clearly underestimates $T_c$, 
as seen dramatically in two dimensions.
On the other hand, the neglect within this theory of transverse 
spin fluctuations must result in an overestimate of $T_{c}$.
It is interesting to note that for hypercubic lattices 
these two effects exactly compensate each other, 
resulting in another accidental coincidence.
The Onsager theory result in Eq. (1) for the transition temperature 
exactly coincides with the N\'{e}el temperature $T_N$ 
for the quantum Heisenberg antiferromagnet on a hypercubic lattice,
as obtained within the Green's function theory,\cite{lines}
and the self-consistent spin fluctuation theory.\cite{neel}

\end{multicols}
\end{document}